\documentclass[conference]{IEEEtran}
\IEEEoverridecommandlockouts

\makeatletter
\def\markboth#1#2{\def\leftmark{\@IEEEcompsoconly{\sffamily}\MakeUppercase{\protect#1}}%
\def\rightmark{\@IEEEcompsoconly{\sffamily}\MakeUppercase{\protect#2}}}
\makeatother

\usepackage[english]{babel}
\usepackage{color}
\usepackage{lipsum}
\usepackage{caption}
\usepackage[noadjust]{cite}
\usepackage[pdftex]{graphicx}
\usepackage{subfig}
\usepackage{caption}
\usepackage{amsmath}
\usepackage{amsfonts}
\usepackage{amssymb}
\usepackage{amsthm}
\usepackage{array}
\usepackage{verbatim}
\usepackage{listings}
\usepackage{algorithm}
\usepackage{algpseudocode}
\usepackage{url}
\usepackage{enumerate}
\usepackage{multirow}

\usepackage{epsfig}
\usepackage{epstopdf}
\usepackage{multicol}
\usepackage[font=footnotesize]{caption}

\usepackage{makecell}
\usepackage{tablefootnote}

\renewcommand{\arraystretch}{2}

\newcommand{\bi}{\begin{itemize}}
\newcommand{\ei}{\end{itemize}}
\newcommand{\be}{\begin{equation}}
\newcommand{\ee}{\end{equation}}

\def\beq{\begin{equation}}
\def\eeq{\end{equation}}
\def\beqa{\begin{eqnarray}}
\def\eeqa{\end{eqnarray}}
\def\beqan{\begin{eqnarray*}}
\def\eeqan{\end{eqnarray*}}

\usepackage{threeparttable}
\usepackage[table,xcdraw]{xcolor}
\usepackage{tabularx}
   \usepackage{multirow}
   \usepackage{booktabs}
   \usepackage{graphicx}
   \usepackage{array, blindtext}
   \usepackage{wrapfig}
\usepackage{pdfpages}
\usepackage{xfrac}
\usepackage{hyperref}

\title{Initial Access in 5G mm-Wave Cellular Networks}
\author{{{\bf Marco Giordani}$^\dagger$, {\bf Marco Mezzavilla}$^\diamond$,  {\bf Michele Zorzi}$^\dagger$ }\\
$^\dagger$ University of Padova, Italy \quad $^\diamond$NYU Wireless, Brooklyn, NY, USA \\
emails: \small{$\{$\texttt{giordani}, \texttt{zorzi}$\}$\texttt{@dei.unipd.it}, \texttt{mezzavilla@nyu.edu}
}}

\begin{document}
\maketitle

\begin{abstract}
The massive amounts of bandwidth available at millimeter-wave frequencies (roughly above $10$ GHz) have the potential to greatly increase the capacity of fifth generation cellular wireless systems.  However, to overcome the high isotropic pathloss experienced at these frequencies, high directionality will be required at both the base station and the mobile user equipment to establish sufficient link budget in wide area networks.  This reliance on directionality has important implications for control layer procedures.  Initial access  in particular can be significantly delayed due to the need for the base station and the user to find the proper alignment for directional transmission and reception.  This paper provides a survey of several recently proposed techniques for this purpose.  A coverage and delay analysis is performed to compare  various techniques including exhaustive  and iterative search, and Context Information based algorithms.  We show that the best strategy depends on the target SNR regime, and provide guidelines to characterize the optimal choice as a function of the system parameters.
\end{abstract}

\IEEEpeerreviewmaketitle

\section{Introduction}
\IEEEPARstart{T}{he} fifth generation (5G) of cellular systems is positioned to address the user demands and business contexts of 2020 and beyond. 
In order to face the continuing growth in subscribers' demand for a better mobile broadband experience, there is a need for significant performance improvements, such as for example\footnote{Nokia White Paper "Looking Ahead to 5G", May 2014. [Online]. Available at \url{http://networks.nokia.com/file/28771/5g-white-paper}}: (i) much greater throughput, with  data rates of  $1$ Gbps or higher, to support ultra-high definition video and virtual reality applications; (ii) much lower latency, less than $1$ ms, to support real time mobile control and Device-to-Device (D2D) applications; (iii) ultra-high reliability and much higher connectivity, to provide seamless service everywhere.

In order to deal with these requirements, some key aspects have been identified to make this future network a reality. Since current microwave ($\mu$-Wave) spectrum under $6$ GHz is  fragmented and crowded,  there has been significant interest in the millimeter-wave (mm-Wave) bands from $10$ GHz to $300$ GHz, where a vast amount of largely unused spectrum is available\footnote{Although strictly speaking mm-Wave bands include frequencies between $30$ and $300$ GHz, industry has loosely defined it to include any frequency  above $10$ GHz. Similarly, we use $\mu$-Wave to denote the traditional cellular bands, i.e., those used by 3G and LTE.}. 
On the one hand, this enormous amount of available spectrum can support the higher data rates required in future mobile broadband access networks. Moreover, the physical size of antennas at mm-Wave frequencies is so small that it becomes practical to build very large antenna arrays (e.g., with $32$ or more elements) to provide further gains from spatial isolation and multiplexing. On the other hand, the increased carrier frequency makes the propagation conditions more demanding than at the lower frequencies traditionally used for wireless services.
For example, blockage becomes an important issue to take care of, as mm-Wave signals, due to their small wavelengths, do not penetrate well most solid materials  (e.g., buildings made of brick) and are subject to very high signal attenuation \cite{key_enabling}.
Another pillar of 5G will be to use many more base stations, deployed according to a heterogeneous network (HetNet) paradigm, combining macro sites with smaller base stations
and using a wide range of radio technologies. These will include LTE, Wi-Fi and any future 5G technologies, integrated flexibly in any combination. 

In this context, the definition of new control layer procedures is critical, in particular initial access (IA), which allows a mobile user equipment (UE) to establish a physical link connection with a base station (BS), a necessary step to access the network. In current LTE systems, IA is performed on omnidirectional channels, whereas beamforming (BF) or other directional transmissions can only be performed after a physical link  is established.  On the other hand, in order to overcome the increased isotropic pathloss experienced at higher frequencies, in 5G mm-Wave cellular systems the IA procedure must provide a mechanism by which the BS and the UE can determine suitable initial directions of transmission. However, directionality can significantly delay the cell search and access procedures, which is a particularly sensitive issue in 5G networks, and thus motivated us to identify and study some performance trade-offs, in terms  of both \emph{delay}, \emph{coverage} and \emph{overhead}.

This work  provides a survey of recent directional IA techniques for mm-Wave cellular systems.  As an extension of our previous work \cite{mine}, in the present paper we compare various search schemes, including exhaustive search, an iterative scheme that successively narrows the search beamwidth, and Context Information (CI) based algorithms, where users are informed about the geolocations of surrounding mm-Wave BSs  through a $\mu$-Wave link. We compare the performance of these approaches in terms  of both misdetection probability and discovery time, under some overhead constraints and as a function of the channel conditions. Our results show that the optimal strategy depends on the target SNR regime, and provide some guidance about the best scheme to use, according to each scenario.


\section{4G-LTE: Initial Access Limitations}
\label{sec:4G_limitations}
In all mobile communication systems, a terminal transitioning from IDLE to CONNECTED mode must perform the following steps: cell search (CS), extraction of system information, and random access (RA). In this section we list the main factors that make 4G-LTE procedures unsuitable for use in a 5G mm-Wave context.


\paragraph{Discovery range mismatch} In LTE systems, acquiring time-frequency domain synchronization during CS is facilitated, as signals are transmitted omnidirectionally in the downlink and BF is used only after a physical link has been established. In mm-Wave bands, it may instead be essential to exploit the BF gains even during the CS phase, since omnidirectional signaling may generate a mismatch between the relatively short range at which a cell can be detected (C-plane range), and the much longer range at which a user could directionally send and receive data (U-plane range) \cite{pa2,zorzi}. 


\paragraph{Multi-connectivity} To ensure sufficient coverage, mm-Wave networks will be much denser. Each user is expected to simultaneously detect multiple potential serving stations, including at least a macro BS operating in the microwave spectrum. Consequently, the IA procedures have to be redesigned in order  to capture this fundamental new feature. We refer to Section \ref{sec:extensions} and \cite{MedHoc2016_MC} for further details.

\paragraph{Deafness and blockage} In mm-Wave cellular networks, IA messages may not be received due to deafness or blockage phenomena. Deafness refers to a situation where the transmit-receive beams do not point to each other, whereas blockage causes a failed message delivery due to a channel drop, which may be related to obstacles, hand rotations, and other mm-Wave-sensitive events. Increasing the transmission power, or waiting for a random back-off time (as done in LTE), are not suitable approaches in mm-Wave networks. Hence, to discriminate among different failed-access causes, new adaptive techniques have to be introduced.

\paragraph{Dynamics-aware access} Due to denser 	topologies, conventional reference signal received power (RSRP)-based association schemes would be highly inefficient in mm-Wave cellular networks, an issue already emerged in HetNets \cite{andrews:load}. However, the challenge with higher frequencies is the need to also account for dynamics such as directionality and intermittency. 

Hence, there is an urge to extend current LTE procedures, and to adapt them to the upcoming mm-Wave related challenges in order to overcome such limitations, or come up with new algorithms and new methods. A natural (and practical) solution is to use beamforming even in the first stages of the initial access procedure, keeping in mind that a fully directional data plane requires a directional IA procedure in the new frequency band.
 On the other hand, when considering an analog multiantenna architecture, directionality means that only one direction can be considered at a time,  thereby losing the broadcast property of the wireless medium, with important  implications for protocol design and delay performance that must be carefully taken into consideration.

The technical issues described in this section call for new initial access procedures and for a detailed assessment of their performance in realistic 5G mm-Wave scenarios. Such a comparison is the main goal of this paper.

\begin{table*}[!t]
\centering
\begin{tabular}{|p{0.25\textwidth} |p{0.37\textwidth}  |  p{0.3\textwidth} | }
\Xhline{2\arrayrulewidth}
\begin{center}
 \textbf{ \large Exhaustive Search}
 \end{center} & \begin{center}
 \textbf{ \large Iterative Search}
 \end{center} & \begin{center}
 \textbf{\large Pure CI-based search}
 \end{center}   \\
\Xhline{2\arrayrulewidth}
 \begin{center}
  {\includegraphics[width=.18\textwidth]{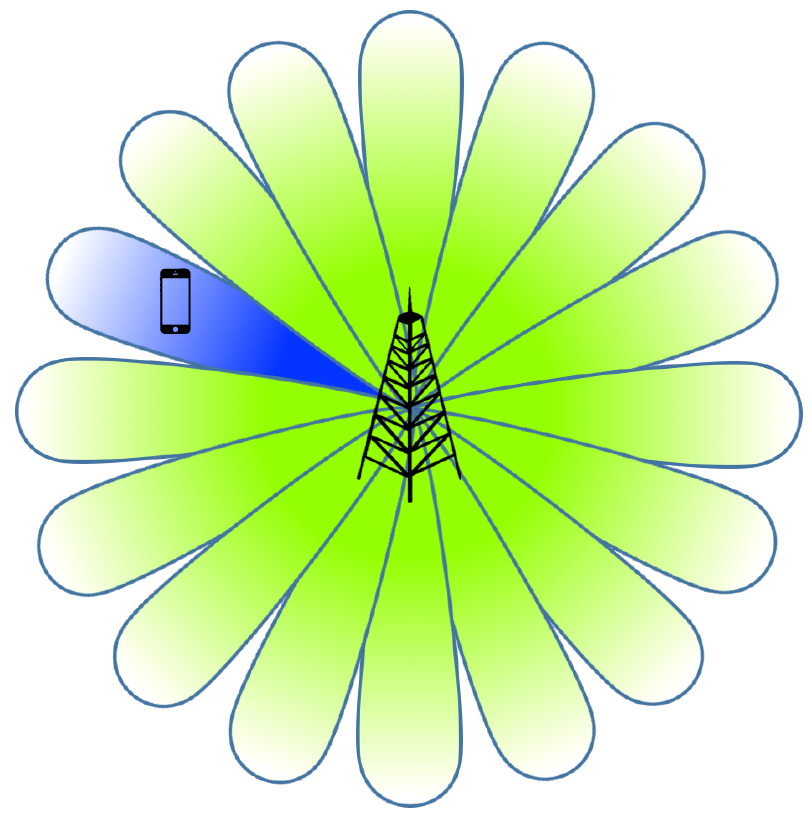} }
  \end{center} &  \begin{center}
  {\includegraphics[width=.18\textwidth]{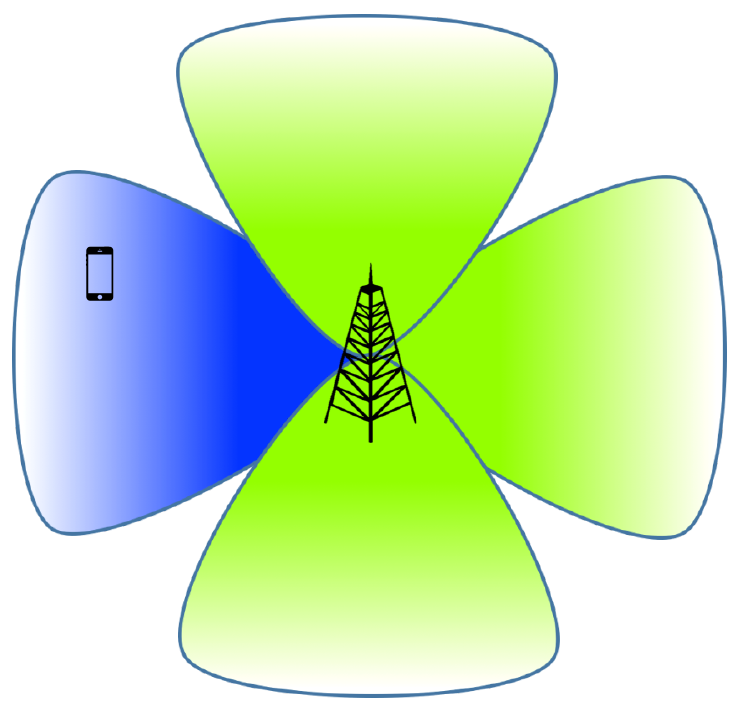} }\quad {\includegraphics[width=.15\textwidth]{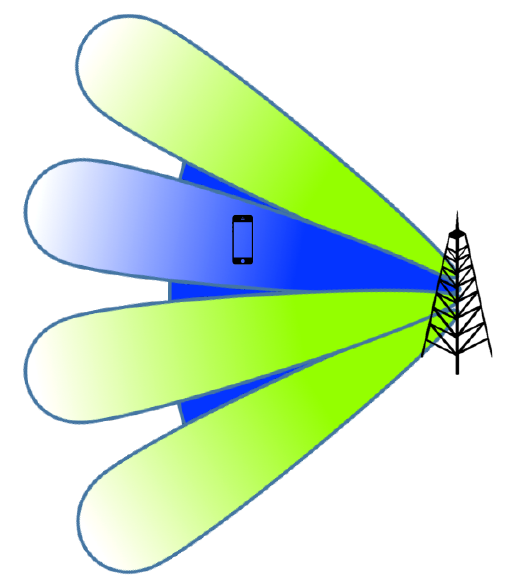} } 
  \end{center} & \begin{center}
  {\includegraphics[width=.3\textwidth]{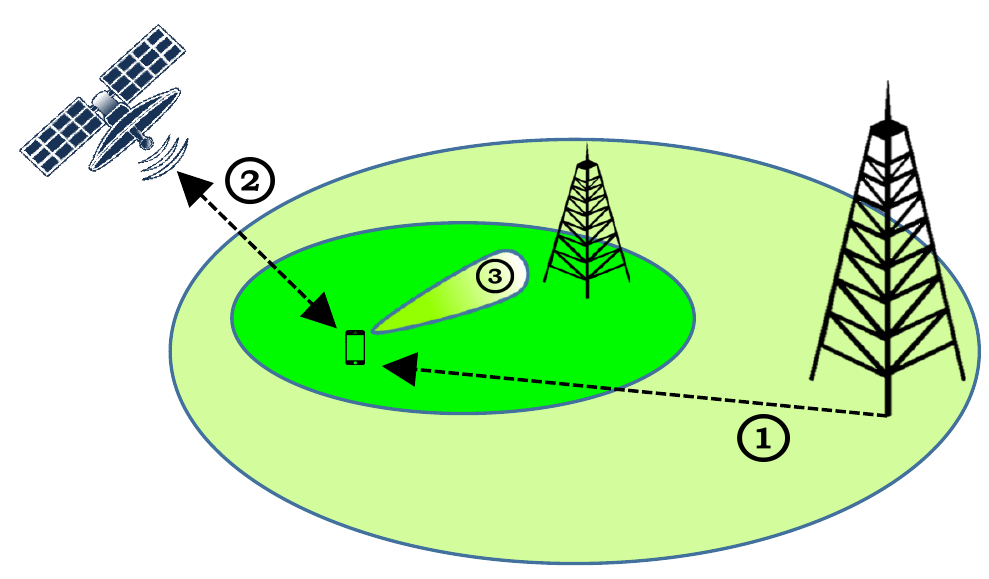} }
  \end{center}  \\
\hline
\begin{itemize}
\item  Good coverage.
\item Suitable for edge users.
\item  High discovery delay.
\end{itemize}
&
\begin{itemize}
\item  Low discovery delay.
\item Not suitable for edge users.
\item   Bad coverage.
\end{itemize}
&
\begin{itemize}
\item  Low discovery delay.
\item   Cost of getting GPS coordinated.
\item  Only LOS scenarios.
\end{itemize}\\
\Xhline{2\arrayrulewidth}
\end{tabular}
\caption{Summary of the three  IA cell search algorithms compared in this work.}
\label{tab:summar}
\end{table*}

\section{Related Work}
\label{sec:rel_work}

Papers  on IA in mm-Wave 5G cellular systems are very recent, since research in this field is just in its infancy. Most literature refers to challenges that have been analyzed in the past at lower frequencies in ad hoc wireless network scenarios or, more recently, in the 60 GHz IEEE 802.11ad WLAN and WPAN scenarios. 

The initial access problem in mm-Wave cellular networks has been considered, for example, in \cite{pa1}, where the authors proposed an exhaustive method to sequentially scan the $360^\circ$ angular space. In \cite{pa2}, a directional cell discovery procedure is proposed,  where base stations periodically transmit synchronization signals, potentially in time-varying random directions, to scan the angular space. Initial access design options are also compared in \cite{pa3}, considering different scanning and signaling procedures, to evaluate access delay and system overhead; the analysis demonstrates significant benefits of low-resolution fully digital architectures in comparison to single stream analog beamforming. Additionally, in order to alleviate the exhaustive search delay issue, paper \cite{pa4} presents a two-phase hierarchical procedure, where a faster user discovery technique is implemented.
%

%
%

On the other hand, Context Information based procedures aim at exploiting knowledge about user and/or BS positions, which are provided by a separate control plane, in order to improve the cell discovery procedure and minimize the delay \cite{pa6}. In \cite{pa7}, booster cells (operating at mm-Waves) are deployed under the coverage of an anchor cell (operating at microwaves). The anchor BS gets control over IA informing the booster BS about user locations, in order to enable mm-Wave cells to directly steer towards the user position. Furthermore, in \cite{pa8}, an evolution of \cite{pa6} is presented, showing how to capture the effects of position inaccuracy and obstacles. Finally, in \cite{Waqas}, the authors study how the performance of analog beamforming degrades in case of angular errors in the available Context Information during the  initial cell search.

In \cite{mine}, we presented a comparison between the exhaustive and the iterative techniques. In this work, we expand the analysis to a CI-based algorithm and describe  a proposed enhancement. 
Our goal is to compare multiple IA procedures under an overhead constraint and to derive the best trade-offs, in terms of both misdetection probability and discovery delay, when considering a realistic dense, urban, multi-path scenario.

\section{Initial Access in  5G mm-Wave Networks}
\label{sec:procedures}

For mm-Wave, we list in the following the initial access steps that enable both the UE and the BS to determine their initial BF directions, in addition to detecting the presence of the BS and of the access request from the UE. 

\subsection{Cell Search for Initial Access}
%
As summarized in Table \ref{tab:summar}, we evaluate three IA-CS  procedures. In this study we focus on \emph{analog} beamforming techniques (considered to be more energy-efficient than their digital counterparts \cite{zorzi}), where each transceiver can look at only one direction per slot. The evaluation of hybrid and fully digital beamforming architectures and a comparison among them are left for future work.

\textbf{Exhaustive search} A brute-force sequential beam searching \cite{pa1}: both users and base stations have a predefined codebook of $N$ directions (each identified by a BF vector) that cover the whole angular space and are used sequentially to transmit/receive. 

\textbf{Iterative search} A two-stage scanning of the angular space \cite{pa4}: in the \emph{first phase}, the BS transmits pilots over wider sectors, while in the \emph{second phase} it refines its search within the best such sector by steering narrower beams. 

\textbf{Context Information based search} This algorithm is articulated into three main stages \cite{pa6}.
\footnote{We note that the procedure we consider in this paper is different from \cite{pa6} in that the CI is available at the UE and concerns the location of the BS. As also done in \cite{Waqas}, this is a more natural way of using CI and assigns the burden of beamscanning to the BS (which would have to do it anyway in the presence of multiple users) rather than to the UE (which can in this case save energy). A more detailed discussion and comparison between the two paradigms is beyond the scope of this paper.}

\begin{enumerate}

\item The macro BS (at $\mu$-Wave frequencies) spreads  omnidirectionally  the GPS coordinates of all the mm-Wave stations within its range.

\item  Each UE gets its own GPS coordinates (this will require a certain energy cost).

\item  According to the information obtained in Steps 1 and 2, each UE geometrically selects the closest BS to connect to and steers a beam towards the direct path. Meanwhile, each mm-Wave BS performs an exhaustive search to detect the best transmit-receive direction.

\end{enumerate}


We refer to \cite{pa1}, \cite{pa4} and \cite{mine} for further procedural details on exhaustive and iterative techniques, and to \cite{pa6,Waqas} for CI-based algorithms.

 In our study, we will focus on the cell search phase described above, which determines whether or not a UE is able to detect a BS and, as will become clear in the following, dominates the overall delay performance.

\subsection{RACHing}
At this stage, the UE does not have any resource or channel available to inform the network about its desire to connect to it;  Random Access provides a means to set up this connection.
Anyway, both the UE and the BS know, from the previous CS phase, the best directions through which they should steer their beams, and therefore they will exchange the following RA messages in one step. The RA is composed of four stages: 1) \emph{RA preamble transmission}, when the UE selects randomly one contention-based signature and sends it to the BS; 2) \emph{RA response}, sent to the UE from the BS and containing an initial timing and power correction as well as some cell specific MAC-layer identifier to uniquely identify the UE in the cell; 3) \emph{connection request message}, sent by the UE desiring initial access that includes, among other data, some authentication and identification information; and 4) \emph{contention resolution phase}, if needed. All subsequent communication can finally occur on scheduled channels.

\section{Performance Evaluation}
\label{sec:sim}
In the simulations, we will assume a static deployment, where no users are moving, so that no handover management or UE motion tracking is required. The parameters are based on realistic system design considerations and are summarized in Table \ref{tab:sim_par}.
To conduct our performance analysis, we assume a slot structure similar to the one described in \cite{pa3}. The primary synchronization signal (PSS) is transmitted periodically, once every $T_{\text{per}}$ seconds, for a duration of $T_{\text{sig}}$ seconds. 

The channel model is based on recent real-world measurements at $28$ GHz in New York City to provide a realistic assessment of mm-Wave micro and picocellular networks in a dense urban deployment. 
Statistical models are derived for key channel parameters including pathloss, number of spatial clusters, angular dispersion and outage.
Further details of this model and its parameters can be found in \cite{mustafa}.


We adopt analog beamforming, implemented through a Uniform Planar Array (UPA), which allows steering towards one direction at a time. 
The total system bandwidth is taken to be $1$ GHz, the transmission powers of BS and UE are set to $30$ dBm and $23$ dBm, respectively, and a noise figure of $5$ dB is assumed. 

In order to detect a PSS signal, the SNR received at the UE has to lie above a certain threshold $\tau$, taken to be $-5$ dB in our results. Decreasing $\tau$ would allow to find more users, at the cost of designing more complex (and expensive) receiving schemes, able to detect the intended signal in more noisy channels.
Each PSS signal has a minimum duration $T_{\text{sig}}=10\:\mu$s, which is deemed to be sufficient to allow proper channel estimation at the receiver.

Simulations are conducted increasing the distance of the UE from the BS. At each iteration, the user is deployed within an annulus having outer radius $R_1$ and inner radius  $R_2 < R_1$, according to a uniform distribution. In order to make reliable measurements through a Montecarlo estimation, each simulation is independently repeated $10^6$ times.

In this study, we evaluate the performance in terms of: (i) \emph{discovery delay}, which is the time required  by the BS and the UE to complete their angular scans (sending and receving PSS control messages, respectively), to determine and select the best directions through which to steer their beams, and (ii) \emph{misdetection probability} (PMD), which is the probability that a UE within the cell is not detected by the BS in the cell search phase, perceiving an SNR below threshold.

\begin{table}[!t]
\renewcommand{\arraystretch}{1.3}
\centering
\begin{tabular}{|c|c|c|}
\hline
\textbf{Parameter} & \textbf{Value} & \textbf{Description}\\
\hline
\hline
 $W_{\rm tot}$ & $1$ GHz & Total system bandwidth\\
\hline
DL  $P_{\rm TX}$ & $30$ dBm & Downlink transmission power \\
\hline
UL  $P_{\rm TX}$ & $23$ dBm & Uplink  transmission power \\
\hline
 NF  & $5$ dB & Noise figure \\
\hline
$f_{\rm c}$ & $28$ GHz & Carrier frequency \\
\hline
min. $\tau$ & $ -5$ dB &  Minimum SNR threshold \\
\hline
  BS antennas & $8 \times 8$  & BS UPA MIMO array size  \\
\hline
UE antennas & $4 \times 4$ or $2 \times 2$& UE UPA MIMO array size\\
\hline
min. $T_{\rm sig}$ & $10 \, \: \mu s$& Minimum signal duration \\
\hline
$\phi_{\rm ov}$ & $5\%$ & Overhead\\
\hline
$T_{\rm per}$ & $T_{\rm sig} / \phi_{\rm ov}$ & Period between transmissions \\
\hline
\end{tabular}
\caption{Simulation parameters.}
\label{tab:sim_par}
\end{table}

\begin{figure}[t!]
\centering
\vspace*{-0.6cm}
 \includegraphics[trim= 0cm 0cm 0cm 0cm , clip=true, width=0.85 \columnwidth]{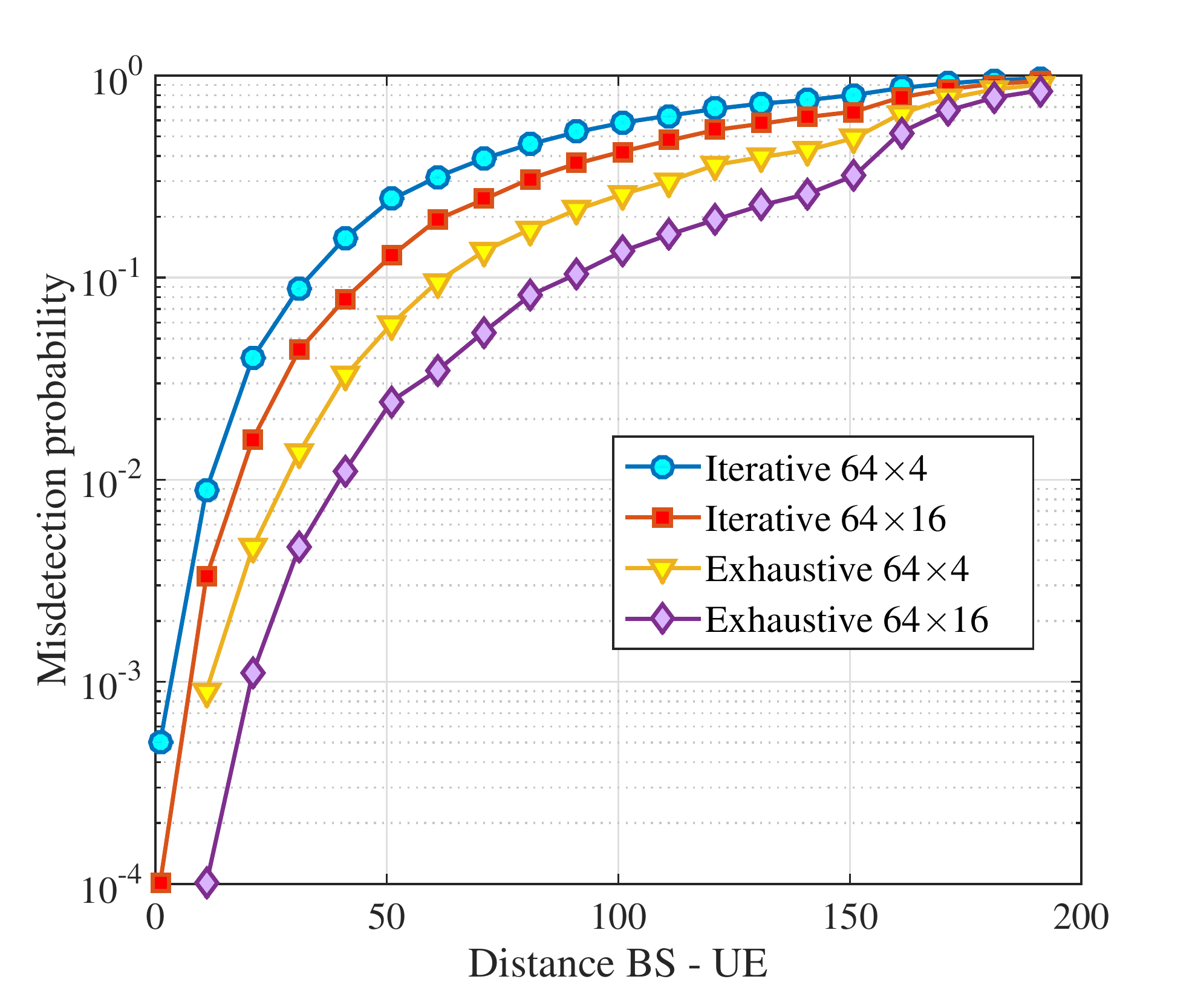}
 \caption{PMD for exhaustive and iterative techniques, when UE receives in  $4$ or $8$ directions, vs. the  BS-UE distance. SNR threshold $\tau=-5$ dB. $T_{\rm sig} = 10 \: \mu$s and $T_{\rm per}=200 \: \mu$s.}
 \label{fig:MDP_CS}
\end{figure}

\begin{figure*}[t!]
\centering
\vspace*{-0.6cm}
\subfloat[][\centering PMD for exhaustive and iterative searches, vs. signal duration $T_{\text{sig}}$. BS-UE distance $d=95$ m.]
   {\includegraphics[width=.42\textwidth]{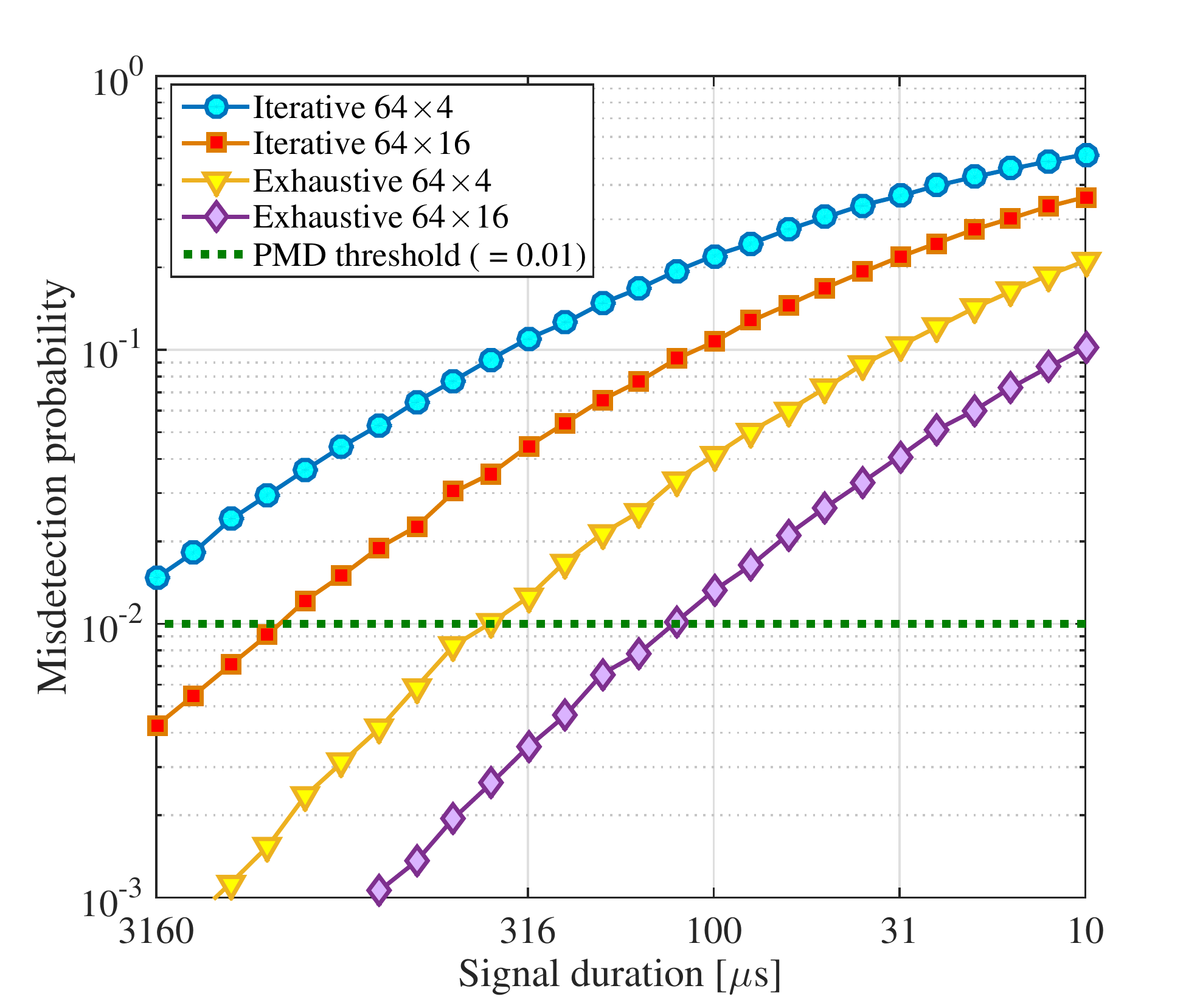}} \quad
\subfloat[][\centering PMD for exhaustive and iterative searches, vs. signal duration $T_{\text{sig}}$. BS-UE distance $d=35$ m.]
   {\includegraphics[width=.42\textwidth]{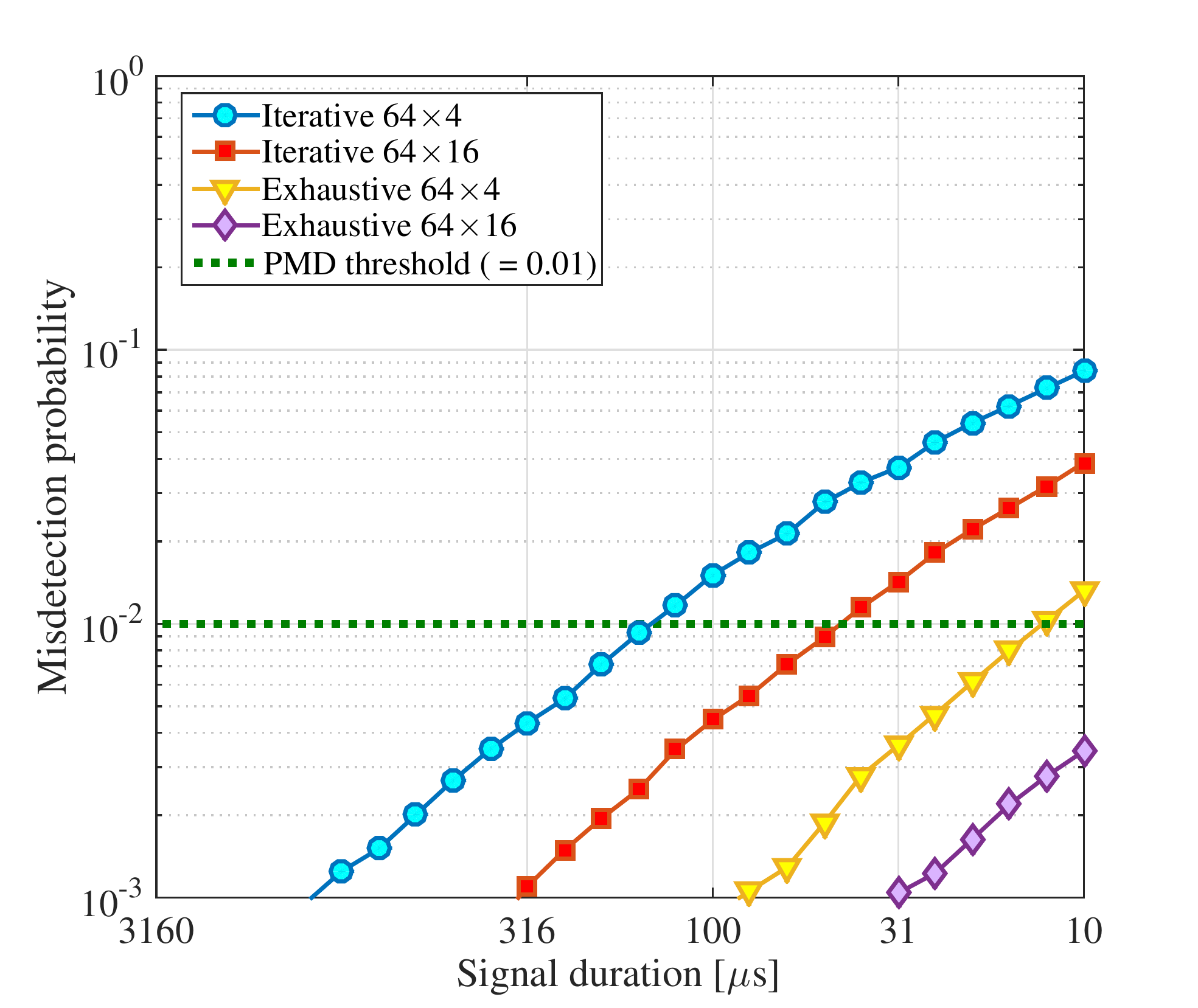}} \\
\caption{Trade-off between delay and PMD.}
\label{fig:slot_dur}
\end{figure*}

In Subsection \ref{sec:exh_vs_it}, we compare two sequential-based IA schemes: the exhaustive search and the iterative technique. The main conclusion of this study is that exhaustive search is likely to be the best IA configuration, when performing CS, especially if we want to provide good coverage with high probability at relatively large distances (e.g., as in the case of edge users in large cells), whereas iterative search may be preferred otherwise (e.g., in case of smaller cells). However, the best technique generally depends on the target SNR and on the considered scenario.

In Subsection \ref{sec:ci_vs_sb}, we compare the CI-based initial access technique with the two sequential schemes. The key finding is that pure CI algorithms may not be suitable for urban scenarios, where links are often NLOS. Nonetheless, we show that more sophisticated directional procedures have the potential to reduce the  discovery delay and grant good coverage.

\subsection{Sequential Approach}
\label{sec:exh_vs_it}

In the exhaustive technique, the BS is equipped with $64$ $(8 \times 8)$ antennas and can steer beams in $N = 16$ directions.
The UE has a set of combining vectors that also cover the whole angular space, and receives PSSs through $4$ wide beams (using only $4$ antennas) or through $8$ narrower beams (using all $16$ antennas). In the iterative first phase, the BS sends PSS messages in $4$ macro directions through $4$ wide beams, using $4$ antennas, while in the second phase it sends the refining PSSs through $4$ narrow beams, through $64$ antennas.

\textbf{Discovery delay:} We consider a minimum signal duration $T_{\text{sig}}=10$ $ \mu$s and a target \emph{overhead} of $\phi_{\text{ov}}=5$ percent. The overhead refers to the percentage of time the channel is used to send digital information across the functional interface between the UE and the BS for the purpose of controlling the IA procedure. The time between two consecutive signal transmissions must be at least $T_{\rm per} = T_{\rm sig} / \phi_{\rm ov} = 200 \: \mu$s. Given that cell search requires $N_s$ slots (one per $T_{\text{per}}$), the discovery delay can be computed as $N_s T_{\rm per} = N_s T_{\text{sig}}/ \phi_{\text{ov}}$. 
The time required for RACHing (the last phase of IA) can be neglected since RA messages are sent through already-set steering directions, without the need to scan again the angular space. Therefore, we argue that the CS latency is the dominant factor when determining the overall IA discovery delay.

Generally, the iterative approach requires fewer slots (and consequently presents a lower discovery delay) with respect to an exhaustive technique, because BSs do not use narrow beams to scan the whole $360^\circ$ angular space, but they just need to refine a macro sector.


\textbf{Misdetection probability:} Since BSs in iterative search transmit over wider beams in the first phase, which results in a reduced beamforming gain, the misdetection probability is higher if compared to exhaustive search, where BSs always use narrower beams that provide higher gains. Moreover, when the UE, in reception mode, exploits just $4$ antennas (receiving PSSs through $4$ beams), the BF gain is reduced as well, also resulting in a higher misdetection probability.

%

\begin{table*}[t!]
\centering
\renewcommand{\arraystretch}{1.6}
\subfloat[][Sequential approach. The minimum signal duration $T_{\rm sig}$ is determined according to Figure \ref{fig:slot_dur}, for each IA scheme.]{
\begin{threeparttable}
\begin{tabular}{|c |c |c| c| c| c|c|c|}
\hline
\begin{tabular}{@{}c @{}} \\ \textbf{Procedure}  \end{tabular} &\begin{tabular}{@{}c @{}}  \\ \textbf{Antennas} \\ \textbf{at the BS } \end{tabular} & \begin{tabular}{@{}c @{}}   \\ \textbf{Antennas} \\ \textbf{at the UE } \end{tabular} & \begin{tabular}{@{}c @{}} \\ $N_{ s}$ \end{tabular}  & \multicolumn{2}{c|}{\textbf{User at $95$} meters} & \multicolumn{2}{c|}{\textbf{User at $35$} meters} \\
\cline{5-8}
& & & & \textbf{Min.} $T_{\rm sig}$ & \textbf{Discovery delay: }$\sfrac{N_s  T_{\text{sig}}}{ \phi_{\text{ov}} }$ & \textbf{Min.} $T_{\rm sig}$ &  \textbf{Discovery delay:} $\sfrac{N_s  T_{\text{sig}}}{ \phi_{\text{ov}} }$ \\
\hline
\hline
Exhaustive $64 \times 4$ & $64$ & $4$ & $80$& $400 \: \mu$s & $640$ ms & $13 \: \mu$s & $20.8$ ms \\
\hline
Exhaustive $64 \times 16$ &  $64$ & $16$ & $144$& $125 \: \mu$s & $360$ ms & $10 \: \mu$s$^*$ & $28.8$ ms \\
\hline
Iterative $64 \times 4$ & \begin{tabular}{@{}c @{}} $4$ in $1^{\text{st}}$ phase \\  $64$ in $2^{\text{nd}}$ phase  \end{tabular} & $4$ & $28$& $> \: 3160 \: \mu$s & $> \: 1760$ ms & $160 \: \mu$s & $89.6$ ms \\
\hline
Iterative $64 \times 16$& \begin{tabular}{@{}c @{}} $4$ in $1^{\text{st}}$ phase \\  $64$ in $2^{\text{nd}}$ phase  \end{tabular} & $16$  & $44$& $ 1580 \: \mu$s & $1390$ ms & $50 \: \mu$s & $44$ ms \\
\hline
\end{tabular}
\begin{tablenotes}
      \scriptsize
      \item $^*$ Minimum allowed signal duration, which is deemed to be sufficient to allow proper channel estimation at the receiver.
    \end{tablenotes}
    \end{threeparttable}
    }\\
    \subfloat[][CI-based approach. The minimum signal duration $T_{\rm sig}$ is determined according to Figure \ref{fig:NCI_vs_CI}, for each CI-based IA scheme.]{
    \begin{threeparttable}
\begin{tabular}{|c |c |c| c| c| c|c|c|}
\hline
\begin{tabular}{@{}c @{}} \\ \textbf{Procedure}  \end{tabular} &\begin{tabular}{@{}c @{}}  \\ \textbf{Antennas} \\ \textbf{at the BS } \end{tabular} & \begin{tabular}{@{}c @{}}   \\ \textbf{Antennas} \\ \textbf{at the UE } \end{tabular} & \begin{tabular}{@{}c @{}} \\ $N_{ s}$ \end{tabular}  & \multicolumn{2}{c|}{\textbf{User at $95$} meters} & \multicolumn{2}{c|}{\textbf{User at $35$} meters} \\
\cline{5-8}
& & & & \textbf{Min.} $T_{\rm sig}$ & \textbf{Discovery delay: }$\sfrac{N_s  T_{\text{sig}}}{ \phi_{\text{ov}} }$ & \textbf{Min.} $T_{\rm sig}$ &  \textbf{Discovery delay:} $\sfrac{N_s  T_{\text{sig}}}{ \phi_{\text{ov}} }$ \\
\hline
\hline
Pure-CI $64 \times 4$ & $64$ & $4$ & $32$& $630 \: \mu$s & $403$ ms & $15 \: \mu$s & $9.6$ ms \\
\hline
Enhanced-CI $64 \times 16$ &  $64$ & $16$ & $64$& $150 \: \mu$s & $192$ ms & $10 \: \mu$s$^*$ & $12.8$ ms \\
\hline
\end{tabular}
\begin{tablenotes}
      \scriptsize
      \item $^*$ Minimum allowed signal duration, which is deemed to be sufficient to allow proper channel estimation at the receiver.
    \end{tablenotes}
    
\end{threeparttable}
}
\caption{Discovery delay that guarantees PMD $< 0.01$ for users at $95$ and $35$ meters from the BS. The table also includes the number of slots ($N_s$) required to implement each IA technique. }
\label{tab:time_all}
\end{table*}

Figure \ref{fig:MDP_CS} shows that in small cells ($0 \div 30$ m range), the misdetection probability is small, and an iterative procedure may be preferred because of its lower delay. Conversely, at mid-range distances iterative methods are not sufficiently reliable and exhaustive techniques would be preferred, whereas for farther users ($100 \div 200$ m range) almost all algorithms perform poorly due to the high probability of channel outage~\cite{mustafa}. The ability of current technologies to adequately support IA only at relatively small distances motivates the search for better IA methods.

\textbf{Trade-off between delay and PMD:} In order to keep the misdetection probability below a certain threshold (i.e., ${ 0.01}$), we can increase the signal duration: if $T_{\text{sig}}$ is increased, the BS transmits its PSS for a longer time in the same sector, so that UEs belonging to that sector can accumulate more energy, resulting in higher SNR and correspondingly reduced PMD. Figure \ref{fig:slot_dur} shows the PMD as a function of $T_{\text{sig}}$ for the different schemes, considering a UE at a distance of $95$ and $35$ meters from the BS, respectively. From these results  we can derive the minimum signal length to meet a certain PMD requirement by reading the abscissa of the intersection of each curve with the horizontal line.



The \emph{discovery delay} in Table \ref{tab:time_all}(a) captures both the required number of slots and the PMD specifications. According to the specific $T_{\text{sig}}$ values obtained from Figure \ref{fig:slot_dur}, the corresponding ${T_{\text{per}}=T_{\text{sig}}/\phi_{\text{ov}}}$ is selected in order to have a constant overhead $\phi_{\text{ov}}=5$ percent.


For example, for the exhaustive $64\times 16$ case in Figure \ref{fig:slot_dur}(a), we see that the corresponding curve intercepts the PMD threshold when the signal duration is around $125\: \mu$s, while, when considering the iterative $64\times 16$ case, a signal duration of around $1580\:\mu$s must be used, to meet the misdetection requirements. Moreover, considering closer users in Figure \ref{fig:slot_dur}(b), we show that exhaustive schemes reach the threshold even when adopting the minimum allowed signal duration $T_{\text{sig}} = 10\: \mu$s. 
Note that in these examples the iterative search, which was the best approach when keeping $T_{\rm sig}$ constant to $10\: \mu$s for all IA schemes, is outperformed by the exhaustive approach if a PMD target is imposed, as its gain in terms of needing fewer slots (roughly a factor of 3 in this example) is outweighed by their longer duration (about one order of magnitude).\footnote{We remark that these computations can also be applied to evaluate the energy consumption performance, although this is not addressed explicitly in this study.}

In general, iterative techniques try to compensate for the lower BF gain in the first phase by collecting enough energy to obtain a sufficiently high SNR, and for this reason require a longer signal duration, despite the lower number of slots required. Conversely, exhaustive searches can operate with shorter slots but need more of them. From the results of Table \ref{tab:time_all}(a), if the goal is to grant good coverage levels to  users at $95$ meters from the BS, an iterative approach is not preferred, as it requires a much longer discovery delay with respect to that of the exhaustive scheme.
On the other hand, if we want to guarantee good PMD just for \emph{closer users} (e.g., in smaller cells), even iterative techniques lead to a sufficiently low discovery delay, despite the longer slots required.

\subsection{CI-Based Initial Access}
\label{sec:ci_vs_sb}

We first consider the pure CI-based technique reported in Table \ref{tab:summar}, which requires $N_s=32$ slots if the BS and the UE are equipped with $64$ and $16$ antennas, respectively (note that in this case the BS has to span the whole angular space, whereas the UE does not need to do any scanning since it knows the BS location and can beamform to it directly). However, if the direct path does not correspond to good channel conditions (e.g., as in NLOS or multi-path scenarios), the beam chosen by the UE may actually be suboptimal. We then propose a more sophisticated scheme, where a successive \emph{beam refinement} is performed: first, the UE points a beam over the direct path inferred via CI. Then, the UE forms additional beams in some adjacent directions, in case they contain a stronger path. 
This requires a total of ${N_s=64}$ slots if three beams are used (the one corresponding to the direct path plus one on each side). 
In both algorithms, we assume that Context Information is not affected by any GPS error and that the time required to collect them is negligible (e.g., the UE already has this information available for some other purposes). 

\begin{figure*}[t!]
\centering
\vspace*{-0.6cm}
\subfloat[][\centering PMD for exhaustive, iterative and CI-based searches, vs. signal duration $T_{\text{sig}}$. BS-UE distance $d=95$ m.]
{ \includegraphics[trim= 0cm 0cm 0cm 0cm , clip=true, width=0.89 \columnwidth]{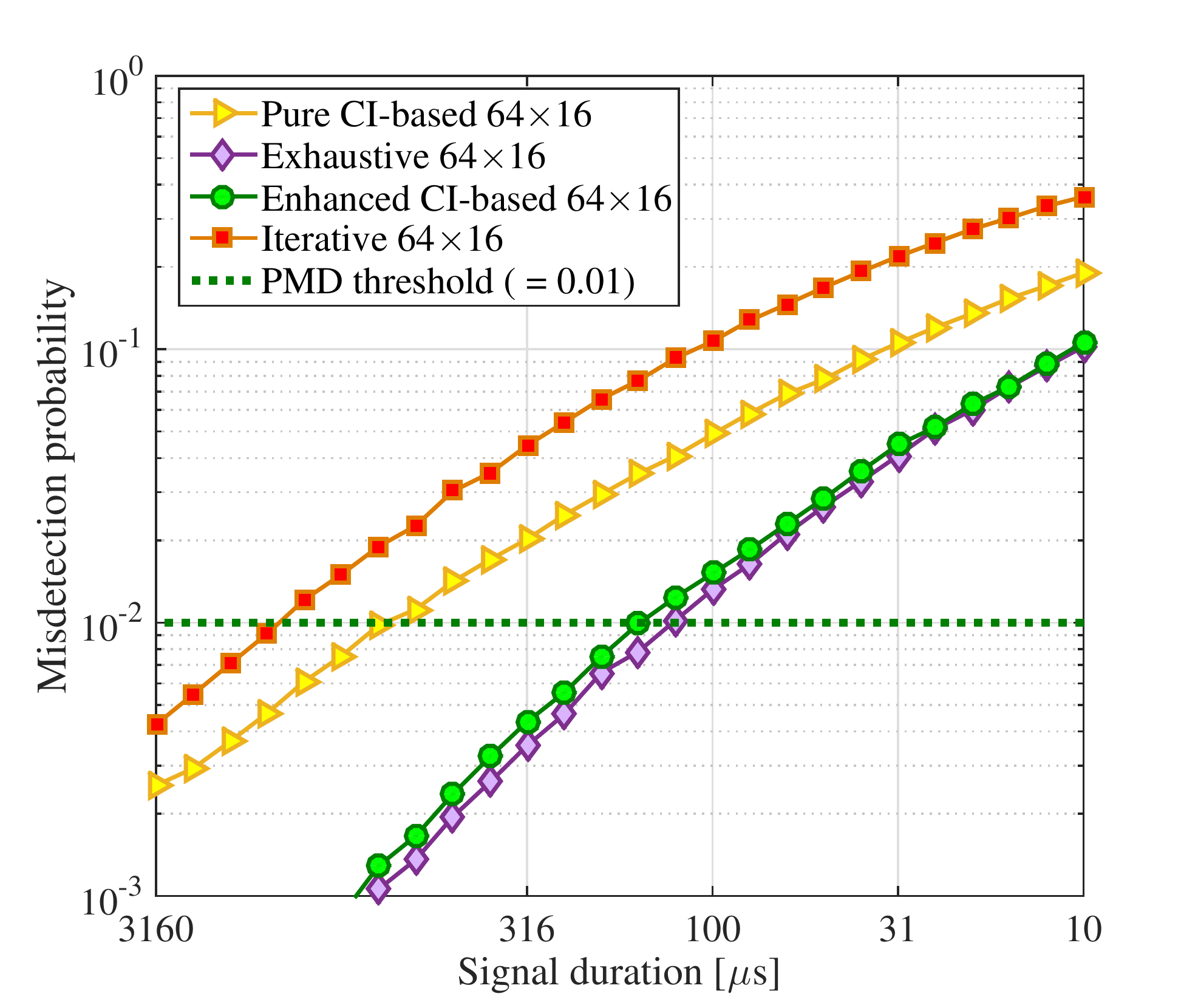}} \quad 
 \subfloat[][\centering PMD for exhaustive, iterative and CI-based searches, vs. signal duration $T_{\text{sig}}$. BS-UE distance $d=35$ m.]
{ \includegraphics[trim= 0cm 0cm 0cm 0cm , clip=true, width=0.89 \columnwidth]{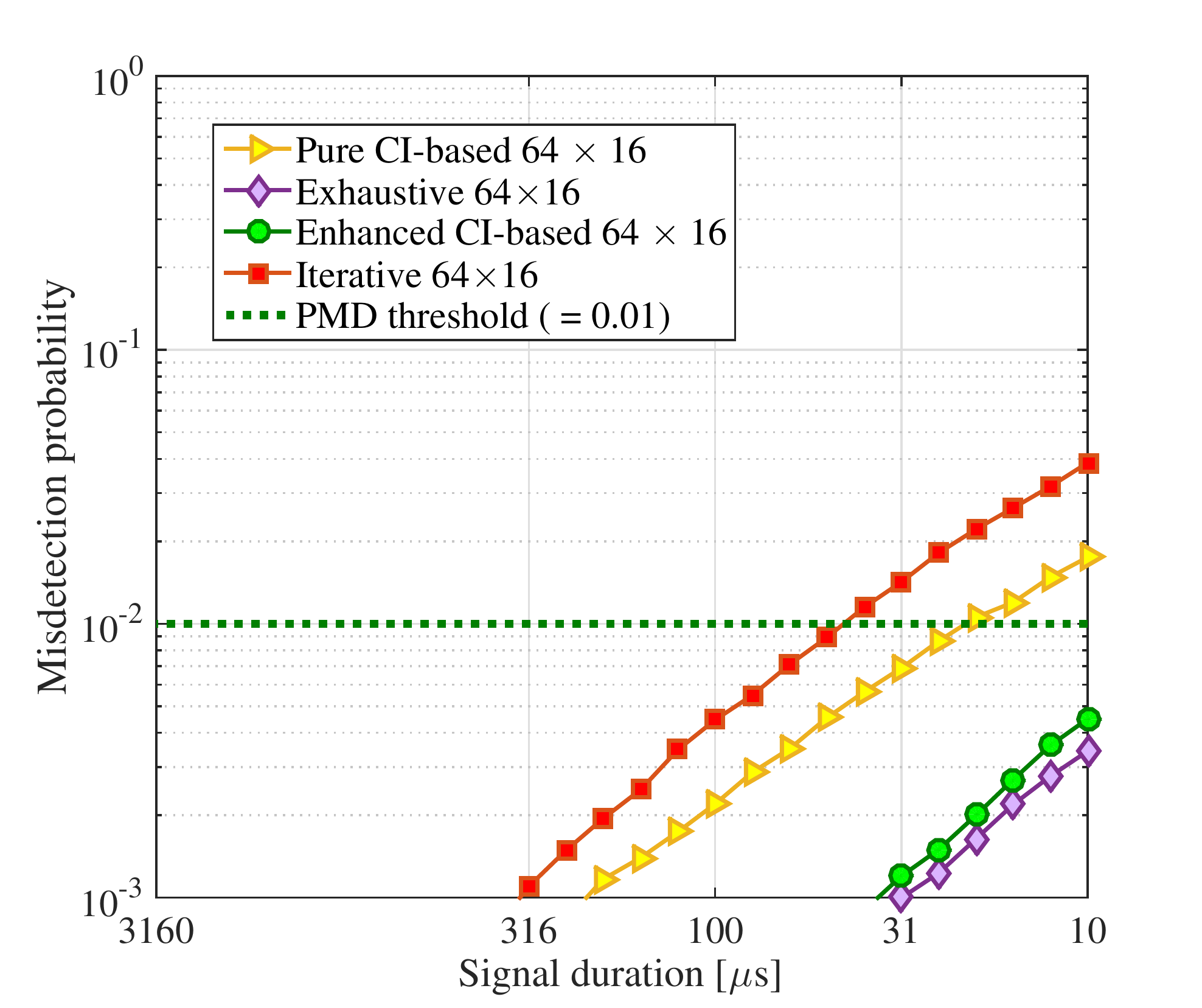}}\\
 \caption{PMD analysis for CI-based initial access.}
 \label{fig:NCI_vs_CI}
\end{figure*}

In Figure \ref{fig:NCI_vs_CI}, we determine the minimum signal duration to meet the usual requirements of PMD ($<0.01$) for  users at $95$ and $35$ m from the BS 
while in Table \ref{tab:time_all}(b) we compute the discovery delay $N_s T_{\rm sig} / \phi_{\rm ov}$, as in the analysis of Subsection \ref{sec:exh_vs_it}.

%

At $95$ m, the discovery delay of the pure CI-based technique is higher than that of the exhaustive approach. 
The reason is that, as mentioned earlier, in a very dense urban environment  the channel propagation is affected by many factors. As a consequence, the direct path obtained through CI-based GPS coordinates may be a poorer choice with respect to what will be selected by the exhaustive procedure in which, instead, the best beam is surely found through a complete scan. 
To meet the PMD requirements, in this case the CI-based algorithm may require very long signals to collect enough energy, with respect to the exhaustive search, thereby greatly increasing the discovery delay. 
On the other hand, the enhanced CI-based technique, where  the refinement is performed, has the potential to drastically reduce the PMD, which almost overlaps with the results obtained through the exhaustive search. 
According to Table \ref{tab:time_all}(b),  the  discovery delay is almost $50$ percent lower than in the exhaustive case.

When considering instead users in good propagation conditions (i.e., at $35$ meters from the BS as in Figure \ref{fig:NCI_vs_CI}(b)), the LOS probability is about $60$ percent (according to \cite{mustafa}), resulting in sufficiently good performance of the pure-CI procedure in terms of  discovery delay (about $9.6$ ms) without any further improvement (which would in turn require to send many more control messages). Therefore, the use of enhanced CI is  justified only to increase the coverage of edge users in large cells.

\section{Initial Access Future Challenges}
\label{sec:extensions}

In this section, we present some important challenges for IA in mm-Wave bands.

\subsection{Beam Tracking}
After one of the IA procedures described in Section \ref{sec:procedures} is performed, the best BS-UE beam pair is determined. 
However, the movement of obstacles and reflectors, or even changes in the orientation of a handset relative to a body or a hand, can cause the channel to rapidly appear or disappear \cite{EW_2016_SNR_tracking}. 
Hence, the selected beam could rapidly change, thus requiring the UE to constantly monitor each potential directional link.
 \emph{Beam tracking} can therefore introduce some latency which lowers the rate at which the network can adapt, and can be a major obstacle in providing robust service in the face of variable link quality.

In \cite{MedHoc2016_MC}, we have proposed a novel multi-cell measurement reporting system to periodically monitor and update the beam direction at both the UE and the BS. A study of the effects of the channel variability and the UE motion over the beam tracking mechanism is left for future work.

\subsection{Multi-Connectivity}
One of the challenges in designing mm-Wave cellular networks is robustness. A likely key feature to meet this requirement is multi-connectivity, where both 5G cells operating at mm-Waves (offering much higher rates) and traditional 4G cells below $6$ GHz (providing much more robust operation) are employed \cite{key_enabling}.

The use of both microwave and
mm-Wave control planes is a key functionality for the IA
technique as well. 
In \cite{MedHoc2016_MC}, we propose a scheme where the UE is  connected to a conventional 4G cell that will perform association and handoff decisions among mm-Wave cells.
Unlike in traditional LTE, the proposed IA system is based on the UE periodically sending uplink (UL) sounding pilots.
We argue that this has several key benefits: (i) the use of UL signals eliminates the need for the UE to send measurement reports back to the network and thereby removes a point of failure in the control signaling path, (ii) if digital beamforming or beamforming with multiple analog streams is available at the mm-Wave cell, then the directional scan time can be dramatically reduced when using UL-based measurements.\footnote{Since the base station is less power constrained than a mobile device, digital or hybrid beamforming will likely be more feasible at the BS side.}

\subsection{Energy-Efficiency}
Periodical signaling for beam tracking, besides increasing the system complexity, results in additional resource and energy consumption. Additionally, GPS coordinates need to be acquired over a certain amount of time when CI-based initial access schemes are performed and, consequently, energy needs to be consumed.
 Finally, the use of digital/hybrid beamforming architectures can reduce the IA discovery time and meet the low latency 5G requirements, but at the expense of requiring much higher energy consumption to feed multiple RF chains. As described in \cite{pa3}, low resolution digital architectures can be a viable solution. In \cite{MedHoc2016_MC}, we show how an UL-based IA framework, where the BS receives sounding reference signals broadcast by the UE, can better exploit the digital BF delay improvements due to less stringent power constraints.

\section{Conclusions and Future Work}
\label{sec:conclusions}

In this work, we have studied, analyzed and compared some possible implementations of initial access techniques for 5G mm-Wave cellular networks, where we argued that directionality should be used also in the initial synchronization-access phase. Our analysis has indeed demonstrated the following key findings:

\begin{itemize}
\item There exists a trade-off between IA delay and misdetection probability: on the one hand,  compared to exhaustive algorithms, iterative techniques require less time to perform the angular search; on the other hand, iterative schemes exhibit higher misdetection probabilities in general, as wider beams provide reduced gains.

\item To guarantee a minimum coverage level at relatively large distances (around $100$ meters from the BS) in realistic channel environments, exhaustive procedures incur smaller  discovery delay and  are therefore to be preferred. Otherwise, for closer users (e.g., in case of ultra-dense cell deployments), iterative techniques still present acceptable delays. 

\item The misdetection probability of pure CI-based approaches is higher than for  exhaustive search, when considering cell-edge users. However, when implementing a simple refinement of the direct link, steering beams through adjacent directions, the PMD drops to acceptable levels, making it possible to  reduce the discovery delay.

\end{itemize}

As part of our future work, the simultaneous steering of narrow beams in multiple directions through  digital and hybrid beamforming architectures will be considered. Furthermore, we will
 study the cost of obtaining CI as well as the performance implications of CI inaccuracy. Finally, we will consider methods that, through historical data about past initial access, can better capture the dynamics of the channel and drive the selection strategy towards more robust cells.

\bibliographystyle{IEEEtran}
\bibliography{bibliography.bib}

\begin{thebibliography}{10}
\providecommand{\url}[1]{#1}
\csname url@samestyle\endcsname
\providecommand{\newblock}{\relax}
\providecommand{\bibinfo}[2]{#2}
\providecommand{\BIBentrySTDinterwordspacing}{\spaceskip=0pt\relax}
\providecommand{\BIBentryALTinterwordstretchfactor}{4}
\providecommand{\BIBentryALTinterwordspacing}{\spaceskip=\fontdimen2\font plus
\BIBentryALTinterwordstretchfactor\fontdimen3\font minus
  \fontdimen4\font\relax}
\providecommand{\BIBforeignlanguage}[2]{{%
\expandafter\ifx\csname l@#1\endcsname\relax
\typeout{** WARNING: IEEEtran.bst: No hyphenation pattern has been}%
\typeout{** loaded for the language `#1'. Using the pattern for}%
\typeout{** the default language instead.}%
\else
\language=\csname l@#1\endcsname
\fi
#2}}
\providecommand{\BIBdecl}{\relax}
\BIBdecl

\bibitem{key_enabling}
L.~Wei, R.~Hu, Y.~Qian, and G.~Wu, ``Key elements to enable millimeter wave
  communications for 5{G} wireless systems,'' \emph{IEEE Wireless
  Communications}, vol.~21, no.~6, pp. 136--143, December 2014.

\bibitem{mine}
M.~Giordani, M.~Mezzavilla, N.~Barati, S.~Rangan, and M.~Zorzi, ``Comparative
  analysis of initial access techniques in 5{G} mm{W}ave cellular networks,''
  in \emph{Proceedings of 50th Annual Conference on Information Sciences and
  Systems (CISS)}, 2016.

\bibitem{pa2}
C.~Barati, S.~Hosseini, S.~Rangan, P.~Liu, T.~Korakis, S.~Panwar, and
  T.~Rappaport, ``Directional cell discovery in millimeter wave cellular
  networks,'' \emph{IEEE Transactions on Wireless Communications}, vol.~14,
  no.~12, pp. 6664--6678, Dec 2015.

\bibitem{zorzi}
H.~Shokri-Ghadikolaei, C.~Fischione, G.~Fodor, P.~Popovski, and M.~Zorzi,
  ``Millimeter wave cellular networks: A {M}{A}{C} layer perspective,''
  \emph{IEEE Transactions on Communications}, vol.~63, no.~10, pp. 3437--3458,
  Oct 2015.

\bibitem{MedHoc2016_MC}
M.~Giordani, M.~Mezzavilla, S.~Rangan, and M.~Zorzi, ``{Multi-Connectivity} in
  {5G} mmwave cellular networks,'' in \emph{2016 15th Annual Mediterranean Ad
  Hoc Networking Workshop (MED-HOC-NET) (Med-Hoc-Net'16)}, Vilanova i la
  Geltru', Barcelona, Spain, Jun. 2016.

\bibitem{andrews:load}
Q.~Ye, B.~Rong, Y.~Chen, M.~Al-Shalash, C.~Caramanis, and J.~Andrews, ``User
  association for load balancing in heterogeneous cellular networks,''
  \emph{IEEE Transactions on Wireless Communications}, vol.~12, no.~6, pp.
  2706--2716, June 2013.

\bibitem{pa1}
C.~Jeong, J.~Park, and H.~Yu, ``Random access in millimeter-wave beamforming
  cellular networks: issues and approaches,'' \emph{IEEE Communications
  Magazine}, vol.~53, no.~1, pp. 180--185, January 2015.

\bibitem{pa3}
\BIBentryALTinterwordspacing
C.~N. Barati, S.~A. Hosseini, M.~Mezzavilla, S.~Rangan, T.~Korakis, S.~S.
  Panwar, and M.~Zorzi, ``Directional initial access for millimeter wave
  cellular systems,'' \emph{CoRR}, vol. abs/1511.06483, 2015. [Online].
  Available: \url{http://arxiv.org/abs/1511.06483}
\BIBentrySTDinterwordspacing

\bibitem{pa4}
V.~Desai, L.~Krzymien, P.~Sartori, W.~Xiao, A.~Soong, and A.~Alkhateeb,
  ``Initial beamforming for mm{W}ave communications,'' in \emph{48th Asilomar
  Conference on Signals, Systems and Computers}, Nov 2014, pp. 1926--1930.

\bibitem{pa6}
A.~Capone, I.~Filippini, and V.~Sciancalepore, ``Context information for fast
  cell discovery in mm-{W}ave 5{G} networks,'' in \emph{Proceedings of 21th
  European Wireless Conference}, May 2015.

\bibitem{pa7}
Q.~Li, H.~Niu, G.~Wu, and R.~Hu, ``Anchor-booster based heterogeneous networks
  with mmwave capable booster cells,'' in \emph{IEEE Globecom Workshops (GC
  Wkshps)}, Dec 2013, pp. 93--98.

\bibitem{pa8}
A.~Capone, I.~Filippini, V.~Sciancalepore, and D.~Tremolada, ``Obstacle
  avoidance cell discovery using mm-{W}aves directive antennas in 5{G}
  networks,'' in \emph{IEEE 26th Annual International Symposium on Personal,
  Indoor, and Mobile Radio Communications (PIMRC)}, Aug 2015, pp. 2349--2353.

\bibitem{Waqas}
W.~B. Abbas and M.~Zorzi, ``Context information based initial cell search for
  millimeter wave 5{G} cellular networks,'' in \emph{Proceedings of 25th
  European Conference on Networks and Communications, EuCNC}, 2016.

\bibitem{mustafa}
M.~Akdeniz, Y.~Liu, M.~Samimi, S.~Sun, S.~Rangan, T.~Rappaport, and E.~Erkip,
  ``Millimeter wave channel modeling and cellular capacity evaluation,''
  \emph{IEEE Journal on Selected Areas in Communications}, vol.~32, no.~6, pp.
  1164--1179, June 2014.

\bibitem{EW_2016_SNR_tracking}
M.~Giordani, M.~Mezzavilla, A.~Dhananjay, S.~Rangan, and M.~Zorzi, ``Channel
  dynamics and {SNR} tracking in millimeter wave cellular systems,'' in
  \emph{European Wireless 2016 (EW2016)}, Oulu, Finland, May 2016.

\end{thebibliography}

\end{document}